# Surface-Enhanced Raman Spectroscopy and Transfer Learning Toward Accurate Reconstruction of the Surgical Zone

Ashutosh Raman[1,2], Ren A. Odion[1,2], Kent Yamamoto[1], Weston Ross[2,4], Tuan Vo-Dinh[1,2,3], Patrick J. Codd[2,4]

[1]*Department of Biomedical Engineering, Duke University, Durham, NC, USA*
[2]*Fitzpatrick Institute for Photonics, Duke University, Durham, NC, USA*
[3]*Department of Chemistry, Duke University, Durham, NC, USA*
[4]*Department of Neurosurgery, Duke University School of Medicine, Durham, NC, USA*
ashutosh.raman@duke.edu

## INTRODUCTION

Raman spectroscopy is a photonic modality defined as the inelastic backscattering of excitation coherent laser light. It is particularly beneficial for rapid tissue diagnosis in sensitive intraoperative environments like those involving the brain, due to its nonionizing potential, point-scanning capability, and highly-specific spectral fingerprint signatures that can characterize tissue pathology [1]. While Raman scattering is an inherently weak process, Surface-Enhanced Raman Spectroscopy (SERS), which is based on the use of metal nanostructure surfaces to amplify Raman signals, has become a compelling method for achieving highly specific Raman spectra with detection sensitivity comparable to conventional modalities such as fluorescence [2]. A unique plasmonics-active nanoplatform, SERS gold nanostars (GNS) have previously been designed in our group to accumulate preferentially in brain tumors [2]. Raman detection, when combined with machine learning and robotics, stands to enhance the diagnosis of surface-level ambiguous tissue during tumor resection surgery, with the potential to improve extent-of-resection and rapidly reconstruct the dynamic surgical field.

Here we demonstrate preliminary results from the use of a SERS-based 2-DOF translational platform to efficiently recreate a tumor embedded in healthy tissue, which is modeled here as a GNS-infused phantom. Transfer learning, specifically through use of the open-source RRUFF mineral database, is employed here to address the dearth of collected biomedical Raman data [3].

## MATERIALS AND METHODS

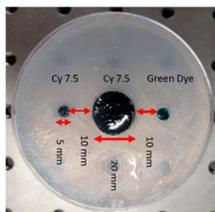

**Fig. 1** Cancer Mimicking Phantom Used in Experiment, with Cy 7.5 regions containing SERS particles

The overall goals of this study were two-fold:
- Combine a Raman system and classification algorithm to discern SERS signal from surrounding tissue-mimicking material, even with changes in material color, and
- Evaluate the ability of a robotic Raman system to reconstruct a sample field rapidly and accurately, with relative dimensions intact.

The experimental setup consists of a portable Raman system, translational stages, and a cancer-mimicking phantom with selectively infused SERS regions.

*A. Phantom Design:* A 3% agarose tissue-mimicking phantom was created for controlled testing of the platform. A 20 mm diameter circular hole was situated in the center of the mold, with two 5 mm diameter holes (fiducials) situated 10 mm edge-to-edge from the middle hole and along the diameter of the overall mold. 1 mL of 1-nM GNS-Cyanine 7.5 (Cy 7.5) solution was mixed with approximately 20 mL of the liquid agarose solution and used to fill the central hole (target), as well as one fiducial. The other fiducial was filled with green tissue dye and no Cy 7.5, to serve as a visually similar control and test the ability of the Raman system to differentiate targets of the same color with and without SERS particles. This is all shown in Fig. 1.

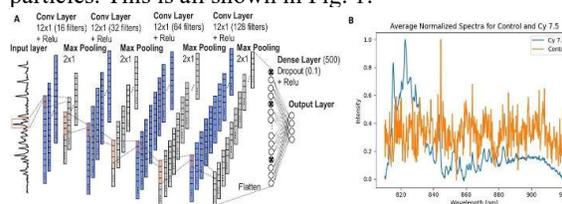

**Fig. 2** A) Neural Network Architecture Adapted from [4] B) Normalized and Iterative Polynomial Fitted Spectral Classes

*B. Neural Network and Transfer Learning:* A convolutional neural network (CNN) was built with the structure shown in Fig. 2A. For the classification of Raman spectra in a biological context, the RRUFF open-source Raman spectra mineral database [3] was first utilized to pre-train the network. Once a validation accuracy of 90% was achieved following classification of the four most prevalent RRUFF mineral classes, all CNN layers were frozen and the final fully-connected layer was replaced with a trainable two-neuron layer.

Using a miniaturized Hamamatsu C13560 Raman system with 15-mW laser power, an integration time of 350 ms, and a sample distance of 10 mm, Raman spectra (n = 100 per class) were collected from solidified agarose and Cy 7.5-infused agarose samples similar in consistency to those used in the experimental phantom. Each spectrum was filtered to only contain intensity values from 810-920 nm, then fit with a 3rd order iterative modified multi-polynomial to remove baseline signal. Then the spectra were min-max normalized, and finally interpolated to 3397 values each, to match the required input feature size for the pre-trained CNN. The average spectra for both classes are shown in Fig. 2B.

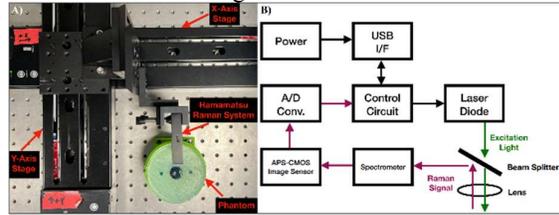

**Fig. 3** A) Experimental Setup Used for Reconstruction and B) Schematic of Hamamatsu Raman System

*C. Translational System Setup:* The Hamamatsu system was attached to two linear translation Stages (Thorlabs LTS300). A 3D-printed mount was used to position the phantom precisely and reproducibly at 10 mm under the laser. The setup is shown in Fig. 3A, with the Hamamatsu Raman system schematic shown in Fig. 3B.

*D. Raster and Reconstruction:* A linear raster within a grid of 30x60 mm was implemented to completely scan the region of interest (ROI) in the phantom, including the tumor-mimicking zone and both fiducials. The stage was moved in 1 mm increments, pausing each time for 350 ms, during which the Raman system automatically captured a spectrum. Each spectrum was then post-processed in the same manner as the training Raman dataset used for CNN fine-tuning. Finally, each spectrum was given a prediction classification of 0 (Control) or 1 (Cy 7.5), and this was assigned to the location from which the sample was taken, to reproduce the ROI. A ground truth reconstruction was also created by manually establishing an origin on the phantom, using the stages to position the laser at the bottom-most edge of the target Cy 7.5 region.

**RESULTS**

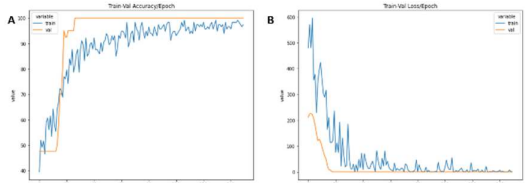

**Fig. 4** A) Accuracy and B) Loss curves for the pre-trained CNN during fine-tuning on Cy 7.5 and Control Agarose

Loss and accuracy curves from fine-tuning are provided in Fig. 4. The CNN had 100% test accuracy without evident overfitting in the binary classification task.

Fig. 5 demonstrates the reconstructed phantom after the raster and prediction. Much like the majority of the surrounding agarose control, the non-Cy 7.5 green dye was not classified as SERS and thus is not visible in the reconstruction. The predicted approximate distance between the two Cy 7.5 samples remains 10mm edge-to-edge, and the general shape of the center sample is preserved at 19mm diameter, while the Cy-7.5 fiducial is slightly misshapen due to misclassification at boundaries. The grid raster lasted 10.2 minutes.

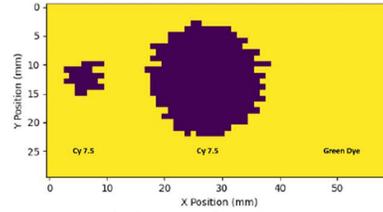

**Fig. 5** Reconstructed Phantom from Raman Raster and CNN Predictions, with Same Orientation as Fig. 1

The ROI area is 1800mm$^2$ and the expected area of the Cy 7.5 regions is 333.8mm$^2$. The predicted area was 328mm$^2$ and thus the accuracy was 98.2%. Comparing the ground truth mask to the predicted reconstruction, the Intersection-over-Union (IoU) score was 84.3%. Comparing boundaries after edge filtering yielded a Boundary IoU of 53%.

**DISCUSSION**

This study achieved high accuracy and IoU, providing a proof-of-concept for a SERS and transfer learning-based pipeline that can aid neurosurgeons seeking rapid noninvasive sensing in vital areas like the brain. This study also shows that SERS does not differentiate between colors, but rather based on presence of GNS agents, which have previously been designed to accumulate in specific brain tumors.

The Boundary IoU was moderate due to the large 1 mm step size, but this can be corrected at the expense of overall raster time; this improved margin detection would make the system more applicable in surgical oncology, primarily in the diagnosis of surface tissue. Furthermore, though the fine-tuning training set was very similar to the phantom, this pipeline still offers value in surgeries for which a biopsy specimen already exists to train CNNs.

Future studies make use of a 6-DOF manipulator to scan on non-flat samples and to simulate system deployment in surgical environments. Studies will also utilize animal tissue to explore effects of tissue structure on classification metrics. In conclusion, the preliminary results support the use of rapid and precise robotic biophotonics to aid intraoperative surface-level diagnosis and margin detection during oncologic surgery.